# Making an analogy between a multi-chain interaction in charge density wave transport and the use wavefunctionals to form soliton-antisoliton pairs

A. W. Beckwith

Department of Physics and Texas Center for Superconductivity and Advanced Materials at the University of Houston
Houston, Texas 77204-5005 USA

## ABSTRACT

A numerical simulation shows that the a massive Schwinger model used to formulate solutions to change density wave (CDW) transport is insufficient for transport of solitons anti-solitons (S-S') through a pinning gap model of CDW transport. However, a model Hamiltonian with Peierls condensation energy used to couple adjacent chains (or transverse wave vectors) permits formation of S-S' that could be used to transport CDW through a potential barrier. There are analogies between this construction and the false vacuum hypothesis used for showing a necessary and sufficient condition in wave functionals for formation of CDW S-S' pairs. This can be established via either use of the Bogomil'nyi inequality or through an experimental artifact resulting from use of .the false vacuum hypothesis to obtain a proportional distance between the S-S' charge centers.

Correspondence: A. W. Beckwith: **projectbeckwith2@yahoo.com**.

PAC numbers: 03.75.Lm, 71.45.Lr, 71.55.-i , 78.20.Ci, 85.25.Cp

# 1. INTRODUCTION

Previously, we demonstrated using integral Bogomol'nyi inequality to present how a soliton-antisoliton (S-S') pair forms.[1,2] We also have shown how wavefunctional formation is congruent with Lin's[3] nucleation of an electron-positron pair as a sufficiency argument to form Gaussian wavefunctionals. Here, we argue our wavefunctional result is equivalent to putting in a multi-chain interaction term in our simulated Hamiltonian system — with a constant term in it proportional to the Peierls gap times a cosine term representing interaction of different CDW chains in our massive Schwinger model.[4] This change of the Hamiltonian term is adding in an additional potential energy term making the problem look like a Josephon junction problem. We found that a single-chain simulation of the S-S' transport problem suffers from two defects. First, it does not answer what are necessary and sufficient conditions for formation of a S-S'. More importantly, numerical simulations of the single-chain transport model demonstrate that barrier penetration requires additional physical conditions. Our numerical simulation of the single-chain problem for CDW involving S-S' gave a resonance condition in transport behavior over time, with no barrier tunneling. The argument we present here is that the false vacuum hypothesis[1,2,4,5] is a necessary condition for the formation of S-S' pairs and that the multi-chain term we add to a massive Schwinger equation for CDW transport is a sufficiency condition for the explicit formation of a S-S' in our CDW transport problem. We begin by a numerical simulation of the single-chain model of CDW, then show how addition of the Peierls condensensation energy permits a S-S' to form. Finally, we explore how this ties in with either the Bogomil'nyi inequality[1,2] and/or the phenomenological Gaussian wave functional model of S-S' pair formation[6] and would permit necessary additional conditions to permit CDW dynamics approaching what we see in the laboratory.

## 2. REVIEW OF THE NUMERICAL BEHAVIOR OF A SINGLE CHAIN FOR CDW DYNAMICS

We are modifying a one-chain model of CDW transport initially pioneered by Dr. John Miller,[7] which furthered Dr. John Bardeen's[8] work on a pinning gap presentation of CDW transport. The single-chain model is a useful way to qualitatively introduce how a threshold electric field initiates transport.[4] However, assuming that the CDW would be easily modeled with a S-S' Gaussian packet, we undertook this investigation to determine necessary and sufficient condition to physically justify use of a S-S' for our wave packet.

We begin using an extended Schwinger model[4] with the Hamiltonian set as

$$H = \int_x \left[ \frac{1}{2 \cdot D} \cdot \Pi_x^2 + \frac{1}{2} \cdot (\partial_x \phi_x)^2 + \frac{1}{2} \cdot \mu_E^2 \cdot (\phi_x - \varphi)^2 + \frac{1}{2} \cdot D \cdot \omega_P^2 \cdot (1 - \cos\phi) \right] \quad (2.1)$$

as well as working with a quantum mechanically based energy

$$E = i\hbar \frac{\partial}{\partial t} \quad (2.2a)$$

and momentum

$$\Pi = (\hbar/i) \cdot \frac{\partial}{\partial \phi(x)} \quad (2.2b)$$

The first case is a one-chain mode situation. Here, $\epsilon \equiv a_D t$ was used explicitly as a driving force, while using the following difference equation due to using the Crank Nickelson[4,9] scheme. We should note that $a_D$ is a driving frequency to this physical system with which we were free to experiment in our simulations. The first index, j, is with regards to 'space', and the second, n, is with regards to 'time' step. Eq. 2.3 is a numerical rendition of the massive Schwinger model plus an interaction term, where one is calling $E = i\hbar \frac{\partial}{\partial t}$ and one is using the following replacement

$$\phi(j,n+1) = \phi(j,n-1) + i \cdot \Delta t \cdot \left( \frac{\hbar}{D} \left[ \frac{\phi(j+1,n) - \phi(j-1,n) - 2 \cdot \phi(j,n) + \phi(j+1,n+1) + \phi(j-1,n+1) - 2\phi(j,n+1)}{(\Delta x)^2} \right] - \frac{2 \cdot V(j,n)}{\hbar} \phi(j,n) \right) \quad (2.3)$$

We use variants of Runge-Kutta in order to obtain a sufficiently large time-step interval so as to be able to finish calculations in a reasonable period of time, while avoiding an observed spectacular blow up of simulated average phase values; which was observed after 100 time-steps at $\Delta t \approx 10^{-13}$. Stable Runge-Kutta simulations require[4] $\Delta t \approx 10^{-19}$:

**[Insert figure 1a about here]**

A second and fully implicit[4] numerical scheme, the Dunford-Frankel, allows us to expand the time step even further. Then, the massive Schwinger model is

$$\phi(j,n+1) = \frac{2 \cdot \tilde{R}}{1 + 2 \cdot \tilde{R}} \cdot (\phi(j-1,n) - \phi(j+1,n)) + \frac{1 - 2 \cdot \tilde{R}}{1 + 2 \cdot \tilde{R}} \cdot \phi(j,n-1) - i \cdot \Delta t \frac{V(j,n)}{\hbar} \phi(j,n) \quad (2.4)$$

**[Insert Fgure 1b about here]**

where $\widetilde{R} = -i \cdot \Delta t \dfrac{h}{2 \cdot D \cdot (\Delta x)^2}$. The advantage of this model is that it is second-order accurate, explicit, and unconditionally stable, so as to avoid numerical blow up behavior. One then gets resonance phenomena as represented by Fig. 1a and Fig. 1b. This is quite unphysical and necessitates making changes.[4]

## 3. AN ADDITIONAL TERM IN THE MASSIVE SCHWINGER EQUATION TO PERMIT FORMATION OF A S-S'

Initially we show how addition of an interaction term between adjacent CDW chains allow a S-S' to form due to the analytical considerations we present here. Finally, we endeavor to show how our interaction-term argument connects with the fate of the false vacuum construction of S-S' terms as previously demonstrated[2,10] when either using the Bogomil'nyi inequality as a necessary condition to the formation of S-S' terms or using the ground-state ansatz argument which still uses the false vacuum hypothesis extensively. Let us now first refer to how we can obtain a soliton via assuming that adjacent CDW terms can interact with each other.

One of our references uses the Bogomil'nyi inequality[2] to obtain a S-S' pair that we approximate via a thin-wall approximation and the nearest-neighbor approximation of how neighboring chains interrelate with one another to obtain a representation of phase evolution as an arctan function w.r.t. space and time variables. Another uses the equivalence of the false vaccum hypothesis with the existence of ground state wavefunctionals in a Gaussian configuration.[6] To whit, either the false vaccum hypothesis itself creates conditions for the necessity of a Gaussian ansatz,[6] else the Bogomil'nyi inequality provides for the necessity of a S-S' pair nucleating via a Gaussian approximation.[2] In our separate model presented herein, however, we find that the interaction of neighboring chains of CDW material permits the existence of S-S' in CDW transport due to the huge $\Delta'$ term added, which lends to a Josephon junction interpretation of this transport problem in CDW dynamics.

Note that in the argument about the formation of a S-S', we use a multi-chain simulation Hamiltonian with Peierls condensation energy[4] to couple adjacent chains (or transverse wave vectors) as represented by

$$H = \sum_n \left[ \frac{\Pi_n^2}{2 \cdot D_1} + E_1[1 - \cos\phi_n] + E_2(\phi_n - \Theta)^2 + \Delta' \cdot [1 - \cos(\phi_n - \phi_{n-1})] \right] \quad (3.1a)$$

with momentum that we define as

$$\Pi_n = (\hbar/i) \cdot \partial/\partial\phi_n \quad (3.1b)$$

We then use a nearest neighbor approximation to use a Lagrangian based calculation of a chain of pendulums coupled by harmonic forces to obtain a differential equation that has a soliton solution. To do this, we write the interaction term in the potential of this problem as

$$\Delta'(1 - \cos[\phi_n - \phi_{n-1}]) \to \frac{\Delta'}{2} \cdot [\phi_n - \phi_{n-1}]^2 + \text{very small H.O.T.s} \quad (3.2)$$

and then consider a nearest neighbor interaction behavior via

$$V_{n.n.}(\phi) \approx E_1[1 - \cos\phi_n] + E_2(\phi_n - \Theta)^2 + \frac{\Delta'}{2} \cdot (\phi_n - \phi_{n-1})^2 \quad (3.3)$$

Here, we set $\Delta' \gg E_1 \gg E_2$, so then

$$V_{n.n.}(\phi)\Big|_{\substack{first \\ order \\ roundoff}} \approx E_1[1 - \cos\phi_n] + \frac{\Delta'}{2} \cdot (\phi_{n+1} - \phi_n)^2 \quad (3.4)$$

that permits us to write

$$U \approx E_1 \cdot \sum_{l=0}^{n+1} [1 - \cos \phi_l] + \frac{\Delta'}{2} \cdot \sum_{l=0}^{n} (\phi_{l+1} - \phi_l)^2 \qquad (3.5)$$

that allowed using $L = T - U$ a Lagrangian based differential equation of

$$\ddot{\phi}_i - \omega_0^2 [(\phi_{i+1} - \phi_i) - (\phi_i - \phi_{i-1})] + \omega_1^2 \sin \phi_i = 0 \qquad (3.6)$$

with

$$\omega_0^2 = \frac{\Delta'}{m_{e^-} l^2} \qquad (3.7)$$

and

$$\omega_1^2 = \frac{E_1}{m_{e^-} l^2} \qquad (3.8)$$

where we assume the chain of pendulums, each of length $l$, leads to a kinetic energy

$$T = \frac{1}{2} \cdot m_{e^-} l^2 \cdot \sum_{j=0}^{n+1} \dot{\phi}_j^2 \qquad (3.9)$$

where we neglect the $E_2$ value. However, having $E_2 \to \varepsilon^+ \approx 0^+$ would tend to lengthen the distance between a S-S' pair nucleating, with a tiny value of $E_2 \to \varepsilon^+ \approx 0^+$ indicating that the distance L between constituents of a S S' pair would get very large.

We did find that it was necessary to have a large $\Delta'$ for helping us obtain a Sine-Gordon equation. This is so if we set the horizontal distance of the pendulums to be $d$, then we have that the chain is of length $L' = (n+1)d$. Then, if mass density is $\rho = m_{e^-}/d$ and we model this problem as a chain of pendulums coupled by harmonic forces, we set an imaginary bar with a quantity $I$ as being the modulus of torsion of the imaginary bar, and $\Delta' = \eta/d$. We have an invariant quantity, which we will designate as: $\omega_0^2 d^2 = \dfrac{1}{\rho \cdot l^2} = v^2$, which, as n approaches infinity, allows us to write a Sine-Gordon equation

$$\frac{\partial^2 \phi(x,t)}{\partial t^2} - v^2 \frac{\partial^2 \phi(x,t)}{\partial x^2} + \omega_1^2 \sin \phi(x,t) = 0 \tag{3.10}$$

with a way to obtain soliton solutions. We introduce dimensionless variables of the form $z = \dfrac{\omega_1}{v} \cdot x$, $\tau = \omega_1 \cdot t$, leading to a dimensionless Sine-Gordon equation we write as:

$$\frac{\partial^2 \phi(z,\tau)}{\partial \tau^2} - \frac{\partial^2 \phi(z,\tau)}{\partial z^2} + \sin \phi(z,\tau) = 0 \tag{3.11}$$

so that

$$\phi_\pm(z,\tau) = 4 \cdot \arctan\left(\exp\left\{\pm \frac{z + \beta \cdot \tau}{\sqrt{1-\beta^2}}\right\}\right) \tag{3.12}$$

where the value of $\phi_\pm(z,\tau)$ is between 0 to $2 \cdot \pi$. As an example of how we can do this value setting, consider if we look at $\phi_+(z,\tau)$ and set $\beta = -.5$. If $\tau = 0$ we can have $\phi_+(z \ll 0, \tau = 0) \approx \varepsilon \approx 0$ and also have $\phi_+(z = 0, \tau = 0) = \pi$, whereas for sufficiently large $z$ we can have $\phi_+(z, \tau = 0) - 2 \cdot \pi$. In a diagram with z as the abscissa and $\phi_+(z,\tau)$ as the ordinate, this propagation of this soliton field from 0 to $2 \cdot \pi$ propagates with increasing time in the positive z direction and with a dimensionless velocity of $\beta$. In terms of the original variables, one has that the 'soliton' so modeled moves with velocity $v \cdot \beta$ in either the positive or negative x direction. One gets a linkage with the original pendulum model linked together by harmonic forces by allowing the pendulum chain as an infinitely long rubber belt whose width is $l$ and which is suspended vertically. What we have described is the flip side of a vertical strip of the belt from $\varphi = 0$ to $\varphi = 2 \cdot \pi$ which moves with a constant velocity along the rubber belt. First, we are using the nearest-neighbor approximation to simplify Eq. 3.4. Then, we are assuming that the contribution to the potential due to the driving force $E_2(\phi_n - \Theta)^2$ is a second order effect. All of this makes for the 'capacitance' effect given by $E_2(\phi_n - \Theta)^2$ not being a decisive influence in deforming the solution, and is a second order effect. This second-order effect contribution is enough to influence the energy band structure the soliton will be tunneling through but is not enough to break up the soliton itself.

## 4. WAVE FUNCTIONAL PROCEDURE USED IN S-S' PAIR NUCLEATION

Traditional current treatments frequently follow the Fermi golden rule for current density[2,11]

$$J \propto W_{LR} = \frac{2 \cdot \pi}{\hbar} \cdot |T_{LR}|^2 \cdot \rho_R(E_R) \tag{4.1}$$

In our prior work we applied either the Bogomil'nyi inequality[1,2,7] or we did more heuristic procedures with Gaussian wave functionals as Gaussian ansatz's to come up with an acceptable wave functional, which will refine I-E curves[2,4] used in density wave transport. For the Bogomol'nyi inequality approach [1,2,7] we modify a de facto 1+1 dimensional problem in condensed matter physics to being one that is quasi-one-dimensional by making the following substitution, namely looking at the Lagrangian density $\varsigma$ to having a time independent behavior denoted by a sudden pop up of a S-S' pair via the substitution of the nucleation pop-up time by

$$\int d\tau \cdot dx \cdot \varsigma \to t_P \cdot \int dx \cdot L \tag{4.2}$$

where $t_P$ is the Planck's time interval. Afterwards, we use the substitution of $\hbar \equiv c \equiv 1$ to write

$$\psi \propto c \cdot \exp\left(-\beta \cdot \int L\, dx\right) \tag{4.4}$$

In the Gaussian wavefunctional ansatz approach,[6] this is more or less assumed as a ground-state energy start to a one-dimensional Hamiltonian of a character which will lead to analytical work in momentum space leading to functional current we derived as being of the form [2,4,6]

$$J \propto T_{if} \tag{4.5}$$

when

$$T_{if} \cong \frac{(\hbar^2 \equiv 1)}{2 \cdot m_e} \int \left( \Psi^*_{initial} \frac{\delta^2 \Psi_{final}}{\delta \phi(x)_2} - \Psi_{final} \frac{\delta^2 \Psi^*_{initial}}{\delta \phi(x)_2} \right) \vartheta(\phi(x) - \phi_0(x)) \wp\, \phi(x) \tag{4.6}$$

where we are interpreting $\wp\, \phi(x)$ to represent taking integration over a variation of paths in the manner of quantum field theory, and $\vartheta(\phi(x) - \phi_0(x))$ is a step-function indicating that we are analyzing how a phase $\phi(x)$ evolves in a pinning gap style potential barrier. We are assuming quantum fluctuations about the optimum configurations of the field $\phi_F$ and $\phi_T$, while $\phi_0(x)$ represents an intermediate field configuration inside the tunnel barrier as we represented by Fig. 3. In both approaches, we pick wavefunctionals with [1,2,6]

$$c_2 \cdot \exp\left(-\alpha_2 \cdot \int d\tilde{x}[\phi_T]^2\right) \equiv \Psi_{final} \tag{4.7}$$

and

$$c_1 \cdot \exp\left(-\alpha_1 \cdot \int dx[\phi_0 - \phi_F]^2\right) \equiv \Psi_{initial} \tag{4.8}$$

with $\phi_0 \equiv \phi_F + \varepsilon^+$ and where $a_2 \cong a_1$. These values for the wavefunctionals appear in the upper right hand side of Fig. 2

**[Place figure 2 about here]**

and represent the decay of the false vacuum hypothesis. This allows us to present a change in energy levels to be inversely proportional to the distance between a S-S' pair [1,2]

$$\alpha_2 \equiv \Delta E_{gap} \equiv \alpha \approx L^{-1} \tag{4.9}$$

We also found that in order to have a Gaussian potential in our wavefunctionals that we needed to have in both interpretations [1,2,6]

$$\frac{(\{\ \})}{2} \equiv \Delta E_{gap} \equiv V_E(\phi_F) - V_E(\phi_T) \tag{4.10}$$

where for the Bogomol'nyi interpretation of this problem we worked with potentials (generalization of the extended Sine Gordon model potential) [1,2]

$$V_E \cong C_1 \cdot (\phi - \phi_0)^2 - 4 \cdot C_2 \cdot \phi \cdot \phi_0 \cdot (\phi - \phi_0)^2 + C_2 \cdot (\phi^2 - \phi_0^2)^2 \tag{4.11}$$

We had a Lagrangian[7] we modified to be (due to the Bogomil'nyi inequality)

$$L_E \geq |Q| + \frac{1}{2} \cdot (\phi_0 - \phi_C)^2 \cdot \{\ \}  \tag{4.12}$$

with topological charge $|Q| \to 0$ and with the Gaussian coefficient found in such a manner as to leave us with wavefunctionals[2] that we generalized for charge density transport. Eq. 4.13 was more or less assumed in the Gaussian wavefunctional ansatz interpretation while we still used Eq. 4.9 and Eq. 4.10 as quasi-experimental imputs into the wavefunctionals according to[1,2,6]

$$\Psi_{i,f}\left[\phi(\mathbf{x})\right]\Big|_{\phi \equiv \phi_{ci,cf}} = c_{i,f} \cdot \exp\left\{-\int d\mathbf{x}\ \alpha \left[\phi_{Ci,f}(\mathbf{x}) - \phi_0(\mathbf{x})\right]^2\right\},  \tag{4.13}$$

In both cases, we find that the coefficient in front of the wavefunctional in Eq. 4.13 is normalized due to error function integration.

## 5: CONCLUSION: SETTING UP THE FRAMEWORK FOR A FIELD THEORETICAL TREATMENT OF TUNNELING.

We have, in the above identified pertinent issues needed to be addressed in an analytical treatment of CDW transport. First, we should try to have a formulation of the problem of tunneling which has some congruence with respect to the Sidney Coleman[5] false vacuum hypothesis. We make this statement based upon the abrupt transitions made in a multi chain model of CDW tunneling that are identical in form to what we would expect in a thin-wall approximation of a boundary between true and false vacuums.[1,2] Secondly, we can say that it is useful to keep a S-S' representation of solutions for charge density transport.[1,2,4].

We explicitly argue that a tunneling Hamiltonian based upon functional integral methods is essential for satisfying necessary conditions for the formation of a S-S' pair.[1,2,4] The Bogomil'nyi inequality[1,2,7] stresses the importance of the relative unimportance of the driving force $E_2 \cdot (\phi_n - \Theta)^2$, which we drop out in our formation of a S-S' in our multi-chain calculation. In addition, we argue those normalization procedures, plus assuming a net average value of the $\Delta'(1 - \cos[\phi_n - \phi_{n-1}]) \to \frac{\Delta'}{2} \cdot [\phi_n - \phi_{n-1}]^2$ + small terms as seen in our analysis of the contribution to the Peierls gap contribution to S-S' pair formation in our Gaussian $\psi \propto c \cdot \exp(-\beta \cdot \int L\, dx)$ representation of how S-S' pairs evolve in a pinning gap transport problem for charge density wave dynamics.[1,2,4,6]

# FIGURE CAPTIONS

Fig. 1a    Beginning of resonance phenomena due to using the traditional Crank-Nickelson numerical iteration scheme of the one chain model.

Fig. 1b    Figure presented completes proof that one chain does not permit tunneling, using Dunford-Frankel numerical scheme for large time stepping.

Fig. 2    Evolution from an initial state $\phi_i$ to a final state $\phi_f$ for a double-well potential (inset) in a quasi 1-D model, showing a kink-antikink pair bounding the nucleated bubble of true vacuum. The shading illustrates quantum fluctuations about the optimum configurations of the field $\phi_F$ and $\phi_T$, while $\phi_0(x)$ represents an intermediate field configuration inside the tunnel barrier. This also shows the direct influence of the Bogomil'nyi inequality in giving a linkage between the distance between constituents of a nucleated pair of S-S' and the $\Delta E$ difference in energy values between $V(\phi_F)$ and $V(\phi_T)$ that allowed us to have a Gaussian representation of evolving nucleated states.

**Figure 1a**
**Beckwith**

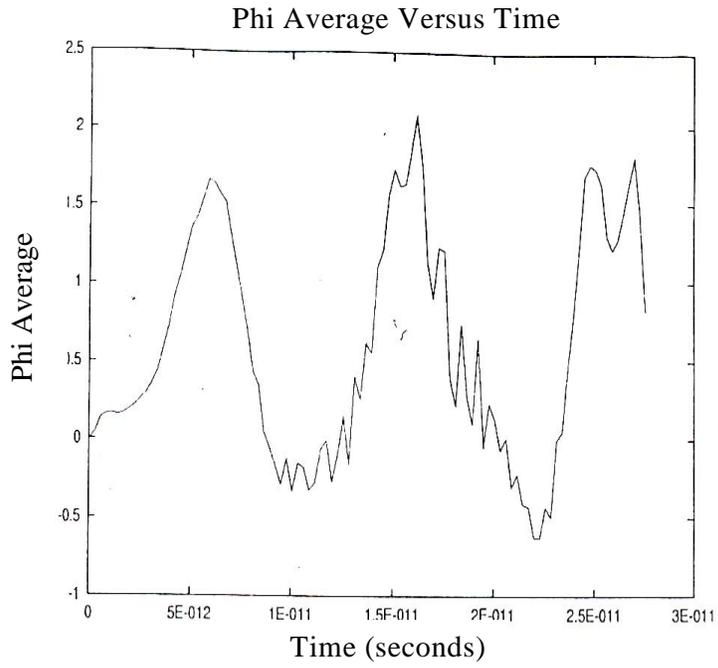

Phi Average Versus Time

**Figure 1b**
**Beckwith**

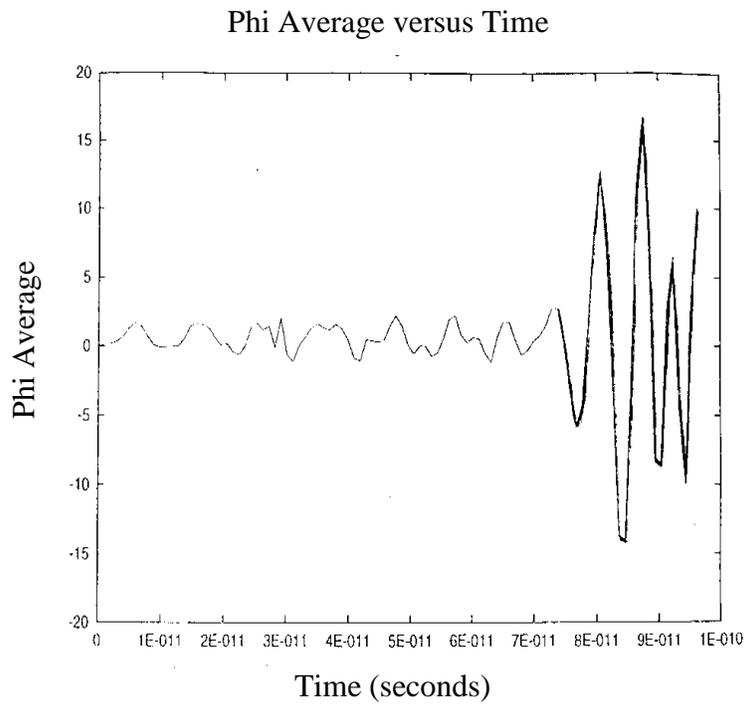

Phi Average versus Time

**Figure 2**
**Beckwith**

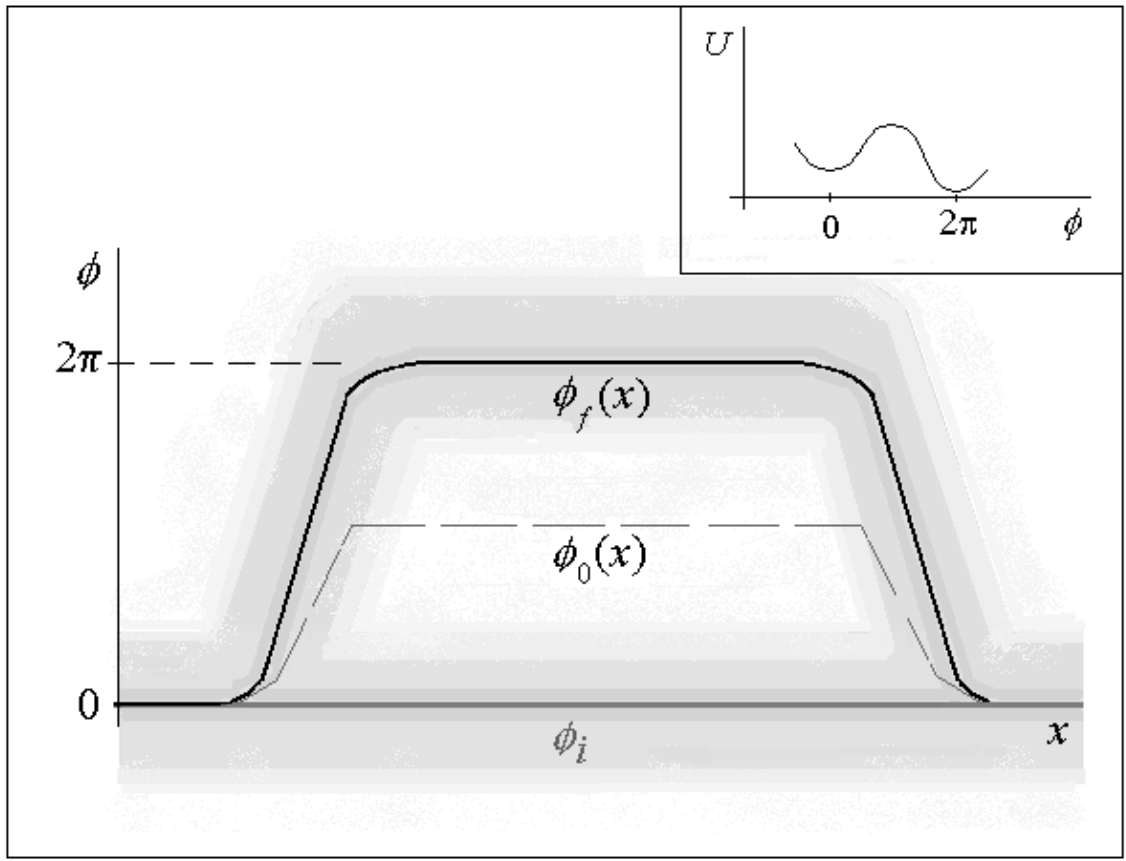